\documentclass[aps,prd,preprint,showpacs,amsmath,amssymb,floatfix]{revtex4}
\usepackage[english]{babel}
\usepackage{epsfig}
\usepackage{enumerate}
\usepackage{subfigure}
\usepackage{rotating}
\usepackage[bookmarks]{hyperref}
\usepackage{color}
\usepackage{multirow} 
\newlength{\www}

\newcommand{\be}{\begin{equation}}
\newcommand{\ee}{\end{equation}}
\newcommand{\ba}{\begin{eqnarray}}
\newcommand{\ea}{\end{eqnarray}}

\newcommand{\bq}{\begin{equation}}
\newcommand{\eq}{\end{equation}}
\newcommand{\bqa}{\begin{eqnarray}}
\newcommand{\eqa}{\end{eqnarray}}
\newcommand{\ben}{\begin{enumerate}}
\newcommand{\een}{\end{enumerate}}
\newcommand{\bc}{\begin{center}}
\newcommand{\ec}{\end{center}}
\newcommand{\bqb}{\begin{eqnarray*}}
\newcommand{\eqb}{\end{eqnarray*}}

\newcommand{\pgsl}{p_g\hskip-0.35cm\slash~}
\newcommand{\qsl}{q\hskip-0.21cm\slash}
\newcommand{\qpsl}{q'\hskip-0.29cm\slash}
\newcommand{\esl}{e\hskip-0.21cm\slash}

\begin{document}

\title{\vspace{1cm}
Associated production of charged Higgs and top at LHC: \\ the role of the
complete electroweak supersymmetric contribution
}
\author{
M.~Beccaria$^{a,b}$,
G.~Macorini$^{a,b}$,
L.~Panizzi$^{c}$,
F.M.~Renard$^e$
and C.~Verzegnassi$^{c, d}$ \\
\vspace{0.4cm}
}

\affiliation{\small
$^a$ $\mbox{Dipartimento di Fisica, Universit\`a del Salento, Italy}$ \\
\vspace{0.2cm}
$^b$ INFN, Sezione di Lecce, Italy\\
\vspace{0.2cm}
$^d$ $\mbox{Dipartimento di Fisica Teorica, Universit\`a di Trieste, Italy}$ \\
\vspace{0.2cm}
$^e$ INFN, Sezione di Trieste, Italy\\
\vspace{0.2cm}
$^f$ Laboratoire de Physique Th\'{e}orique et Astroparticules,
Universit\'{e} Montpellier II, France
}

\begin{abstract}
The process of charged Higgs production in association with a top quark at the LHC has been calculated
at the complete NLO  electroweak level both in a Two Higgs Doublets Model and in the Minimal Supersymmetric
Standard Model, assuming a mSUGRA breaking scheme.
We have numerically explored the size of the one-loop corrections  in two
typical supersymmetric scenarios, with particular attention to the $\tan\beta$ dependence,
and we have found that they
remain perturbatively small but possibly sizable, reaching a 20\% limit for extreme values of $\tan\beta$,
when the complete set of Feynman diagrams is taken into account.

\end{abstract}

% \pacs{12.15.Lk,13.75.Cs,14.80.Ly,14.80.Cp}

%\begin{flushright}
% PTA/07-47\\
%FNT/T 2007/02
%\end{flushright}

\maketitle

\section{Introduction}
\label{sec:intro}

The processes of production of a charged Higgs boson will be extensively
exploited to search for new physics beyond the standard model at the LHC.
Most extensions of the Standard Model (SM), such as two-Higgs-doublet models (2HDM) or the
Minimal Supersymmetric Standard Model (MSSM), enlarge the minimal SM Higgs
sector predicting the existence of charged Higgs particle(s).
Since the discovery of a charged Higgs boson would be a distinctive signature of new physics,
an exhaustive comprehension of its production mechanism appears to be mandatory.

Depending on the charged Higgs boson mass, different production mechanisms are dominant:
if $m_{H^+} < m_t - m_b$ the main source of charged Higgs is the $t \bar{t}$ production and the
subsequent decay of the top $t \to H^+ b$~\cite{Kunszt:1991qe},
while for a heavier charged Higgs boson the dominant process is the associated production with
heavy quarks~\cite{Barger:1993th}-\cite{Dittmaier:2009np}.
Also the associated production with $W$ gauge boson has been analysed \cite{Dicus:1989vf}-\cite{Hollik:2001hy}, but this
process is suppressed with respect to the other two mechanisms of production.

We will focus our analysis on the associated production with a top quark,
which is also an important mechanism of top production and should be considered
in the analysis of single top production at the LHC \cite{Bernreuther:2008ju}.
At the lowest perturbative order,
it is well known that this process is particularly sensitive to the value of the parameter $\tan\beta$,
{\it i.e.} the ratio of the neutral Higgs vacuum expectation values $v_2/v_1$, which appears in the Yukawa coupling $tbH$.
Being proportional to $m_b \tan\beta$, the coupling is enhanced for large values of
$\tan\beta$
% and resummation techniques should be used to get rid of such large contributions~\cite{Carena:1999py}.
and this enhancement allows a direct check of the 2HDM structure of the model, not necessarily involving SUSY.
Supersymmetric corrections, on the other hand, can be only investigated looking at the loop structure of the process.\\

Due to its relevance, the process of production of charged Higgs in
association with a top quark has been extensively studied at higher orders and many important results have been obtained.
The NLO corrections in QCD and SUSY QCD 
in the five-flavour scheme ({\it i.e.} including the bottom quark as a parton of the sea)
have been computed in Refs.~\cite{Zhu:2001nt}-\cite{Berger:2003sm}
and the same corrections in the four-flavour scheme, together with a comparison
of the results in the two schemes have been computed in Ref.~\cite{Dittmaier:2009np}.
As a general feature, while QCD corrections are generally found to be large and positive
and nearly independent of $\tan\beta$,
SUSY corrections appear to be sizable and negative for large $\tan\beta$.
For what concerns the NLO electroweak (EW) contribution,
the subset of Yukawa SUSY EW corrections has been computed in Refs.~\cite{Jin:2000vx}-\cite{Belyaev:2002eq} in both
the five-
and four-flavour schemes: all of these papers assume that the Yukawa part of the correction is the leading one
and they get some large one-loop contributions for ``extreme'' values of $\tan\beta$.

Given the possible relevance of the considered process, which might require a more accurate prediction, we have performed
in this paper a complete NLO MSSM EW calculation. This includes all the EW diagrams that were neglected
in the previous analyses and also the total QED radiation, that has never been computed for this process and whose
effects might be {\it a priori} relevant,
as we know from previous recent calculations of our group~\cite{Beccaria:2007tc,Beccaria:2008av}.

The paper is organized as follows. Sect.~\ref{sec:kin} will be
devoted to a description of the shape and of the basic properties of the
parton level amplitudes for $bg\to t H^-$ at Born and at one-loop
level. A rigorous treatment of QED radiation has been performed to obtain reliable values for the observables and will
be described in Sect.~\ref{subsection:qed}.
In Sect.~\ref{sec:oneloop} the numerical one-loop effects
on the production rates and distributions for a couple of meaningful
SUSY benchmark points will be shown, together with a discussion of the results.

\section{Kinematics and Amplitudes of the process $b~g\to tH^-$}
\label{sec:kin}

\subsection{Kinematics}

The kinematics of the process $b~g\to tH^-$ 
is expressed in terms of the $b$ quark momentum $p_b$, helicity $\lambda_b$, spinor $u(p_b,\lambda_b)$,
the $t$ quark momentum $p_t$, helicity $\lambda_t$, spinor $\bar u(p_{t},\lambda_{t})$, with:
\bq
p_b=(E_b;0,0,p)~~~~~~~~
~p_{t}=(E_{t};p'\sin\theta,0,p'\cos\theta)~,
\eq
\noindent
the gluon momentum $p_g$, helicity $\lambda_g$, polarization vector $e_g$ and the Higgs boson momentum $p_H$:
\bq
p_g=(p;0,0,-p)~~~~~~~~e_g(\lambda_g)=(0;{\lambda_g\over\sqrt{2}},-~{i\over\sqrt{2}},0)
\eq
\bq
~p_{H}=(E_{H};-p'\sin\theta,0,-p'\cos\theta)
\eq
We also use the s-channel and u-channel momenta:
\bq
q=p_g+p_b=p_{H}+p_{t}~~~s=q^2~~
~~~q'=p_{t}-p_g=p_b-p_{H}~~~u=q^{'2}
\eq

The invariant amplitude of the process $b~g\to tH^-$ will be decomposed on a set of 8 forms $J_{k\eta}$,
where $\eta$ represents the chirality $R,L$ (sometimes denoted $\eta=+1,-1$). The 8 scalar functions $N_{k\eta}(s,t,u)$
will be computed in the next subsection from the various Born and one loop diagrams. 
\bq
A=\sum_{k} J_{k\eta} N_{k\eta}(s,t,u)
\eq
\bq
J_{1\eta}=\pgsl\esl P_{\eta}
~~~~~~
J_{2\eta}=(e.p_{t}) P_{\eta}
\eq
\bq
J_{3\eta}=\esl P_{\eta}
~~~~~~
J_{4\eta}=(e.p_{t})\pgsl P_{\eta}
\eq
with $P_{\eta}=P_{R,L}=(1\pm\gamma^5)/2$.

The 8 helicity amplitudes $F_{\lambda_b,\lambda_g,\lambda_t}$
are obtained from Dirac decompositions  of the 8 invariant forms.

Averaging over initial spins and colours and summing over final spins and colours
with
\bq
\sum_{col}<{\lambda^l\over2}><{\lambda^l\over2}>=4
\eq
one gets the elementary cross section:
\bq
{d\sigma\over d\cos\theta}= {\beta'\over768\pi s\beta}
\sum_{spins}|F_{\lambda_b,\lambda_g,\lambda_t}|^2
\eq
where $\beta=2p/\sqrt{s}$, $\beta'=2p'/\sqrt{s}$. \\

\subsection{Born and one loop amplitudes}

\begin{figure}
\centering
\begin{minipage}{0.4\textwidth}
\vskip -18pt\epsfig{file=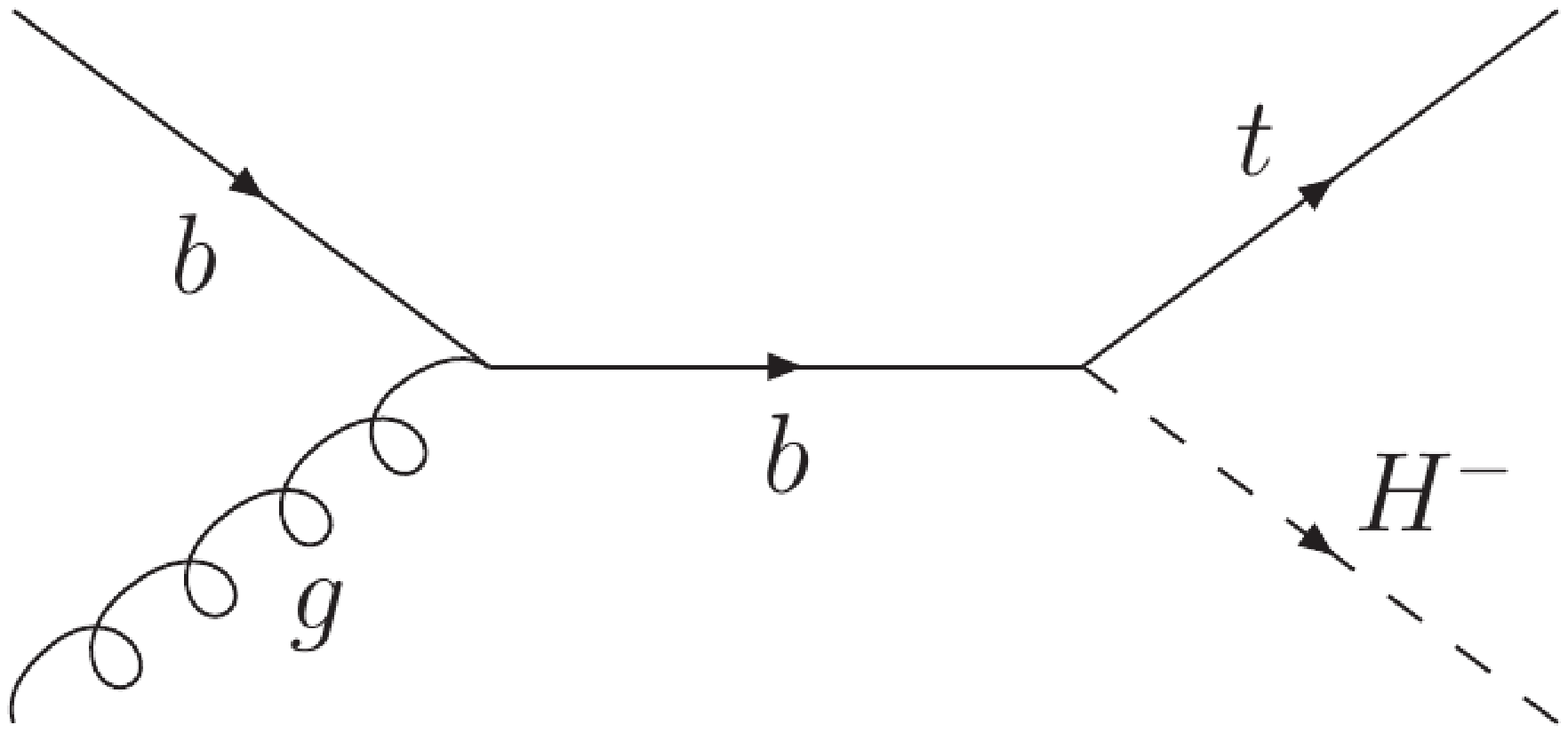, width=\textwidth, angle=0}
\end{minipage}
\hskip 50pt
\begin{minipage}{0.4\textwidth}
\epsfig{file=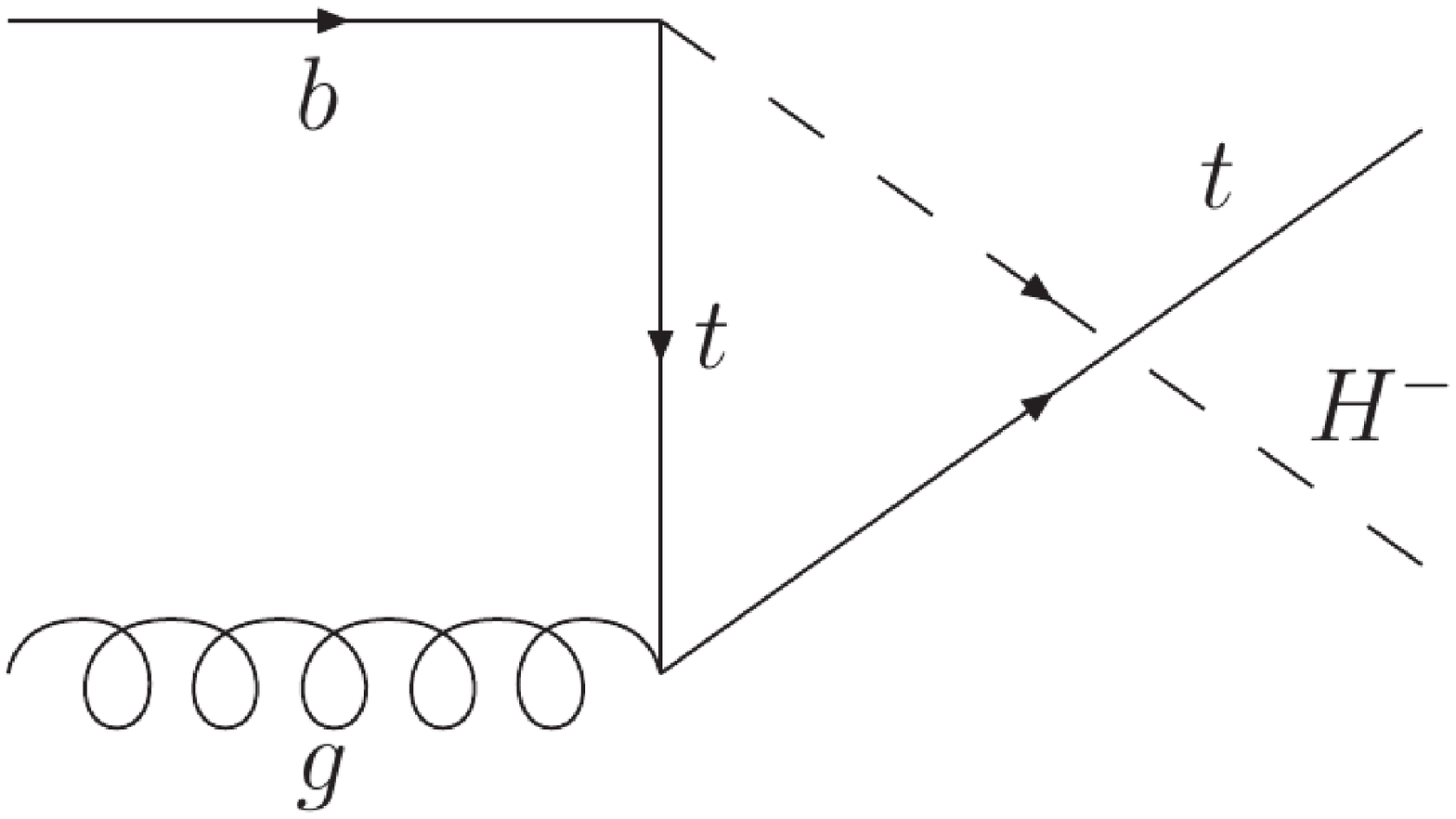, width=\textwidth, angle=0}
\end{minipage}
\caption{Born diagrams: s-channel bottom exchange and u-channel top exchange.}
\label{fig:Born}
\end{figure}

The Born terms result from the s-channel $b$ exchange
and the u-channel $t$ exchange of Fig.~\ref{fig:Born}:

\bq
A^{Born~s}=-({g_s\over s-m^2_b})({\lambda^l\over2})
\bar u(t)
[c^L(b\to t H^-)P_L+c^{R}(b\to t H^-)P_R]
(\qsl+m_b)\esl u(b)
\eq
leads to the scalar function
\bq
N^{Born~s}_{1~\eta}=-g_s({\lambda^l\over2})
{c^{\eta}(b\to t H^-)\over s-m^2_b}
\eq
and
\bq
A^{Born~u}=-({g_s\over u-m^2_{t}})
({\lambda^l\over2})\bar u(t)\esl(\qpsl+m_t)
[c^L(b\to t H^-)P_L+c^{R}(b\to t H^-)P_R]u(b)
\eq
to
\bq
N^{Born~u}_{1\eta}=-g_s({\lambda^l\over2})
{c^{\eta}(b\to t H^-)\over u-m^2_{t}}
\eq
\bq
N^{Born~u}_{2\eta}=-2g_s({\lambda^l\over2})
{c^{\eta}(b\to t H^-)\over u-m^2_{t}}
\eq
with the $btH^-$ couplings
\bq
c^{L}(b\to t H^-)= {em_t\cot\beta\over\sqrt{2}s_WM_W}~~~~c^{R}(b\to t H^-)= {em_b\tan\beta\over\sqrt{2}s_WM_W} 
\eq

\noindent The one loop EW terms can be classified as:
\begin{itemize}
  \item[---] counter terms for $b,t, H^-$ lines and $btH^-$ coupling constants. We follow the on-shell scheme
%   \cite{onshell}
  in which all of the counter terms can be computed in terms of self-energy diagrams. For what concerns the 
  $H^-$ line and the $btH^-$ coupling, we use the procedure given in \cite{Wan:2001rt} which takes into
  account the $G^-,H^-$ mixing and expresses the counter term for $\tan\beta$ in terms of $W-H$ mixing self-energy.
  Other procedures, e.g. \cite{Eberl:2001eu} or \cite{Freitas:2002pe},
  would lead to a similar divergence cancellation but 
  differ by minor finite contributions.
  \item[---] self-energy corrections for internal $b$ and $t$
  propagators;
  \item[---] s-channel left triangles:
  $(Vqq)$, $(Sqq)$, $(\chi \tilde{q}\tilde{q})$ and right triangles:
  $(qSV)$, $(Vq' q)$, $(qVS)$, $(Sf'f)$, $(fSS')$;
  \item[---] u-channel up triangles: $(ffV)$, $(ffS)$, $(SSf)$;
  and down triangles: $(VSq)$, $(q'qV)$, $(SVq)$, $(f'fS)$, $(S'Sf)$;
  \item[---] direct boxes: $(qqq'V)$, $(qqq'S)$, $(\tilde{q}\tilde{q}\tilde{q'}\chi^-_j)$;
  crossed boxes: $(qqVS)$, $(qqSV)$, $(\tilde{q}\tilde{q}\chi_i\chi_j)$, $(qqSS')$;
   twisted boxes: $(qqq'S)$, $(qqq'V)$, $(\tilde{q}\tilde{q}\tilde{q'}\chi_j)$.
\end{itemize}

All these contributions have been computed using the
usual decomposition in  terms of Passarino-Veltman functions and 
the complete amplitude has been implemented in the numerical code \verb+PumaMC+.\\

We have checked the cancellation of the UV divergences
among counter terms, self-energies and triangles.
This cancellation occurs separately inside 8 sectors, i.e. s-left L or R, s-right L or R,
u-up L or R, u-down L or R.\\

Another useful check can be done using the high energy behaviour of the amplitudes.
High energy rules \cite{Beccaria:2002cf,Beccaria:2003yn} predict the logarithmic
behaviour of these amplitudes at one loop level.
They use splitting functions for external particles
$b,t,H$ and Renormalization Group effects on the parameters
appearing in the Born terms.

By using the logarithmic expansions of the Passarino-Veltman functions \cite{Beccaria:2007ni}
we have checked that the amplitudes obtained by summing
the contributions of the above self-energy, triangle and box
diagrams satisfy these rules. 

At high energy we first observe the mass suppression of $N^{Born~s+u}_3$  as well as
the cancellation of $N^{Born~s+u}_1$.  One remains with only $N^{Born~u}_2$
and the 2 helicity amplitudes $F_{-++,+--}$:
\bqa
F^{Born}_{-,+,+}&\to&
-~{eg_sm_t \cot\beta\over s_WM_W}
({\lambda^l\over2})\cos{\theta\over2}
({1-\cos\theta\over1+\cos\theta})
\eqa
\bqa
F^{Born}_{+,-,-}&\to&
-~{eg_sm_b \tan\beta\over s_WM_W}
({\lambda^l\over2})\cos{\theta\over2}
({1-\cos\theta\over1+\cos\theta})
\eqa
At one loop logarithmic level the aforementioned rules \cite{Beccaria:2002cf,Beccaria:2003yn} predict
the corrections:
\bqa
F_{-,+,+}&=&
F^{Born}_{-,+,+}\{1+{1\over2}[c(b\bar b ~L)
+c(t\bar t ~R)]+c^{ew}_{-,+,+}(H^-)\}
\eqa
\bqa
F_{+,-,-}&=&
F^{Born}_{+,-,-}\{1+{1\over2}[c(b\bar b ~R)
+c(t\bar t ~L))]+c^{ew}_{+,-,-}(H^-)\}
\eqa
\noindent
in which $c(b\bar b ~L)$, $c(t\bar t ~R)$ represent the $b,t$ splitting functions
and $c^{ew}_{\mp,\pm,\pm}(H^-)$ the total of the $H^-$ splitting and of the parameter renormalization
of the $btH^-$ couplings through
$\delta g/g -\delta M/M_W +\delta m_t/m_t
-\delta\tan\beta/\tan\beta$ and
$\delta g/g -\delta M_W/M_W+\delta m_b/m_b 
+\delta\tan\beta/\tan\beta$.

The result being:
\bqa
F_{-,+,+}&=&F^{Born}_{-,+,+}\{1+[{\alpha\over4\pi}]\{
-[{1\over3c^2_W}]\log^2\frac{s}{m^2_Z}
-[{1\over9c^2_W}]\log^2\frac{-t}{m^2_W}
\nonumber\\
&&
+~{1-4c^2_W\over12s^2_W c^2_W}[\log^2\frac{-u}{M_Z^2}]
-~~{1\over2s^2_W}[\log^2\frac{-u}{M_W^2}]\}~\}
\eqa 

\bqa
F_{+,-,-}&=&F^{Born}_{+,-,-}\{
1+[{\alpha\over4\pi}]\{
-[{1+2c^2_W\over12s^2_Wc^2_W}]\log^2\frac{s}{m^2_Z}
-[{1\over2s^2_W}]\log^2\frac{s}{m^2_W}
\nonumber\\
&&+{1\over18c^2_W}[\log^2\frac{-t}{M_W^2}]
-~{1\over6s^2_W}[\log^2\frac{-u}{M_W^2}]
~\}~\}
\eqa 
\noindent
with the absence of linear logarithmic terms as noticed in \cite{Beccaria:2009vj} .

Taking our complete one loop computation and retaining only the logarithmic
parts of the B,C,D Passarino-Veltman functions appearing in the various diagrams, we do
recover the above expressions for the 2 leading amplitudes.\\

\subsection{QED radiation}
\label{subsection:qed}

The computation of the real photon radiation contributions has been
performed according to Ref.~\cite{Beccaria:2008jq}.  The matrix element
has been calculated analytically with the help 
of \verb+FeynArts+~\cite{Kublbeck:1990xc}
and  \verb+FormCalc+~\cite{Hahn:1998yk}. Infrared (IR) singularities
have been regularized within mass regularization, {\it i.e.} giving a
small mass to the photon, and the phase space
integration has been performed using the phase space slicing method.

Concerning the choice of the parton distribution functions (PDFs) and
their factorization, we follow Ref.~\cite{Beccaria:2008jq}. 
The PDFs used through this computation are the 
LO QCD parton distribution functions
CTEQ6L~\cite{Pumplin:2002vw}.
The factorization of the bottom PDF has been performed in the
$\overline{\mbox{MS}}$ scheme at the scale  $Q =
(m_t+m_{H^{-}})$. If the DIS factorization
scheme is used,
the differences in the numerical value of the one-loop EW
effects are of the order of 0.01\% in all the considered mSUGRA benchmark points.

\begin{figure}
\centering
\epsfig{file=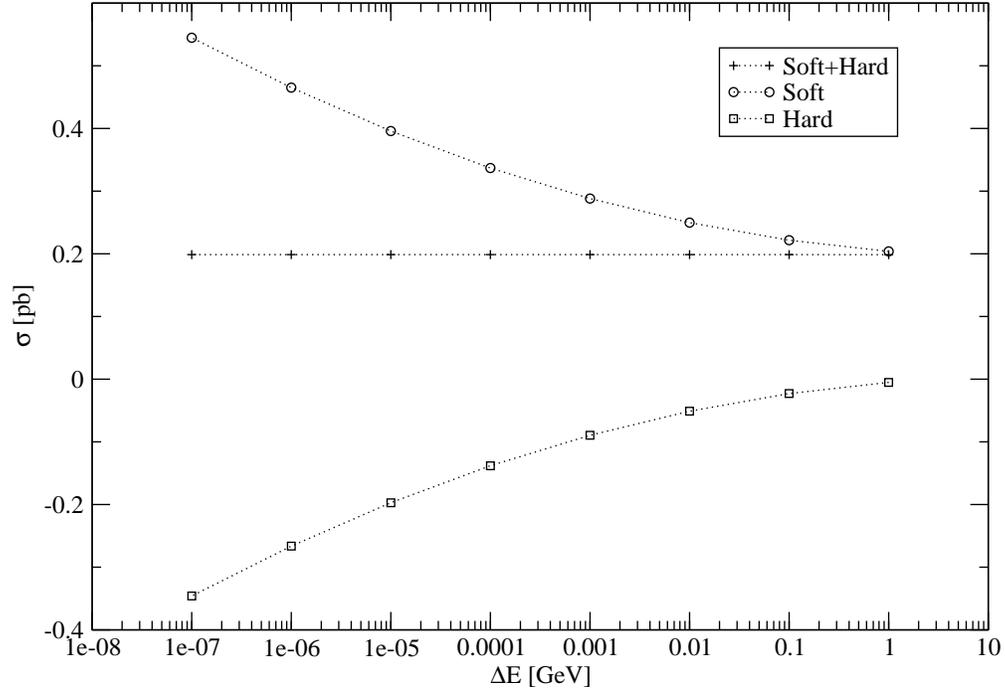, width=0.8\textwidth,
  angle=0}\\[50pt]
\epsfig{file=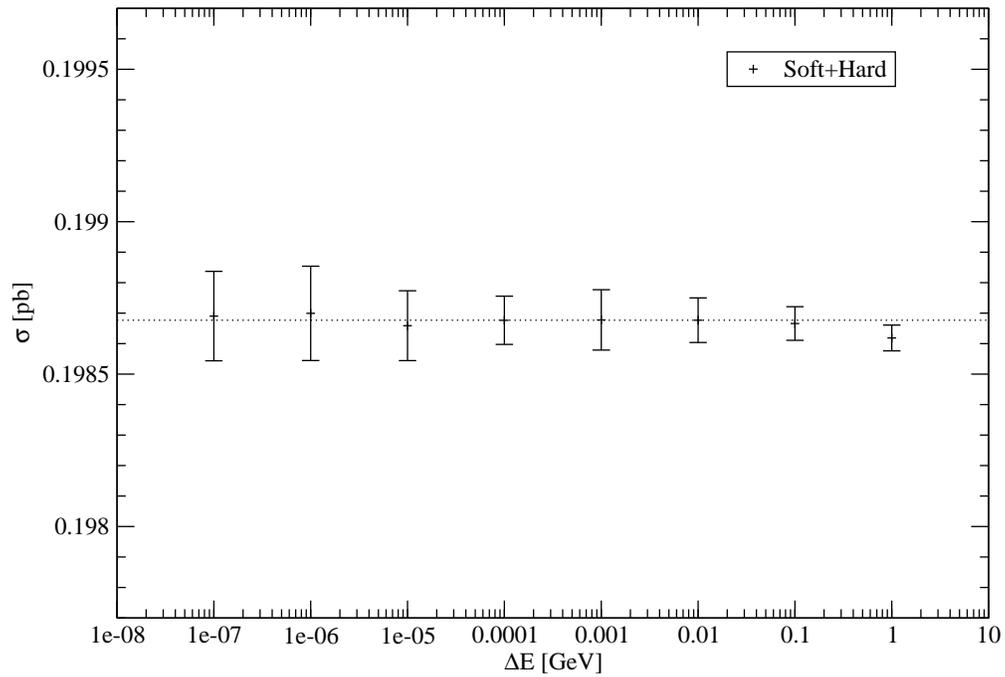, width=0.8\textwidth, angle=0}
\caption{Upper panel: dependence of the ${\cal O}(\alpha)$ 
soft plus virtual and hard cross sections on the soft-hard 
separator $\Delta E$. Lower panel: independence of the sum of ${\cal O}(\alpha)$ soft plus 
virtual and hard cross sections 
of the separator $\Delta E$.}
\label{fig:QEDcheck}
\end{figure}

The phase space slicing method introduces a fictitious separator $\Delta E$ in
the integration over the photon energy. 
As a check of our computations we have verified that, for sufficiently small $\Delta E$ values,
the final cross section is independent on the choice of $\Delta E$.
Despite  the strong sensitivity to $\Delta E$ of the soft 
and of the hard 
cross section, {\it c.f.} the upper panel of Fig.~\ref{fig:QEDcheck}, the
dependence of the total result on $\Delta E$ is far below
the integration uncertainties (lower panel of Fig.~\ref{fig:QEDcheck}).

\section{One Loop Results}
\label{sec:oneloop}

For the numerical evaluation of the one-loop corrections we have considered as SM inputs the values in Tab.\ref{tab:SMinputs}.
The strong coupling constant has been evaluated at the
renormalization scale $Q = m_t+m_{H^-}$ and its numerical value will be given below.
Since we have performed our computations in the on-shell scheme, we have
evaluated the pole mass of the bottom quark starting from the $\overline{MS}$ mass at NLO in QCD, obtaining $m_b=4.58$~GeV.

\begin{table}[htbp]
\begin{tabular}{@{\extracolsep{5pt}}|c|c|}
  \hline
  Coupling constants & $\alpha=1/137.035999 \quad \alpha_s(M_Z)=0.118$ \\
  \hline
  Gauge boson masses & $\quad M_W = 80.424~\rm{GeV} \quad M_Z = 91.1876~\rm{GeV}$ \\
  \hline
  Quark Masses       & $\begin{array}{lll} m_u = 47~\rm{MeV} & m_c = 1.55~\rm{GeV} & m_t = 170.9~\rm{GeV}\\
                                           m_d = 47~\rm{MeV} & m_s = 0.15~\rm{GeV} & \overline{m_b}(\overline{m_b}) = 4.2~\rm{GeV}
                        \end{array}$ \\
  \hline
  Lepton Masses      & $\begin{array}{ccc} m_e = 0.51099906~\rm{MeV} &
                                           m_\mu = 105.6583~\rm{MeV} &
                                           m_\tau = 1.777~\rm{GeV} \end{array}$ \\
  \hline
\end{tabular}
\caption{Numerical values of SM inputs}
\label{tab:SMinputs}
\end{table}

As a first step, we have analysed the distributions of the invariant mass of the final states $d\sigma/dM_{inv}$
and the total cross section for a couple of representative SUSY benchmark points
(assuming a mSUGRA supersymmetry breaking): LS2 \cite{Beccaria:2006ir} and SPS1a \cite{SPS}.
The characteristics of the benchmark points, together with the mass of the charged Higgs $H^-$
and the value of $\alpha_s(Q)$, are shown in Tab.~\ref{tab:bench}.
The two benchmarks are characterized by largely different input parameters at GUT scale, leading to different scenarios
for low energy spectra: the LS2 point is an optimistic ``light SUSY'' scenario, while the SPS1a point is a standard and widely
studied scenario for phenomenological analyses with higher masses. Moreover, the two points differ for the $\tan\beta$ values:
LS2 features a very large $\tan\beta = 50$, while in SPS1a $\tan\beta = 10$.
The complete spectra at low energy have been obtained running the parameters
through the code \verb#SUSPECT#\cite{Djouadi:2002ze}. The values of $\tan\beta$ at low energy
have been translated from those obatained in the  $\overline{DR}$ scheme used by \verb#SUSPECT# to the
values in the on-shell scheme through the relation $\tan\beta(OS) = \tan\beta(\overline{DR}) - \delta\tan\beta(OS)|_{\rm{finite}}$. The values we have obtained for LS2 and SPS1a are 60.5 and 10.4 respectively.

\begin{table}[htbp]
 \begin{tabular}{|c|ccccc||c|c|c|}
 \hline
 ~mSUGRA scenario~ & $\quad m_0 \quad$ & $\quad m_{1/2} \quad$ & $\quad A_0 \quad$ & $\quad \tan\beta \quad$ & $\quad \textrm{sign } \mu \quad$ & $\quad H^- \quad$ & $\quad \alpha_s(Q) \quad$  \\
 \hline
 LS2         & 300 & 150 & -500 & 50 & + & 229.6 & ~0.0965325~ \\
 SPS1a       & 100 & 250 & -100 & 10 & + & 412.1 & ~0.0922963~ \\
 \hline
 \end{tabular}
\caption{input parameters for the mSUGRA benchmark points and mass of the charged Higgs $H^-$ (all values with mass dimension are in GeV)}
\label{tab:bench}
\end{table}

The resulting total cross sections and K-factors (where, as usual, $K = \sigma_{1-loop}/\sigma_{Born}$)
are shown in Tab.\ref{tab:sigmaK}.
Due to the very mild dependence of our calculations on the PDF factorization scheme,
only the results obtained in the $\overline{MS}$ scheme are shown.
We have performed the analysis considering both the whole supersymmetric spectra
(labelled ``SUSY'' in the following discussion)
and the ``SUSY constrained'' two-Higgs-doublet-model (2HDM) scenarios
obtained from the original spectra considering only loops involving Higgs bosons and SM particles
({\it i.e.} without charginos, neutralinos and sfermions).

\begin{table}[htbp]
\begin{tabular}{|c|c|cc|cc|}
\hline
\multirow{2}{*}{~mSUGRA scenario~} & \multirow{2}{*}{$\quad \sigma_{Born} \quad$} &
\multicolumn{2}{c|}{SUSY} & \multicolumn{2}{c|}{2HDM} \\
\cline{3-6}
 & & $\quad \sigma_{1-loop} \quad$ & ~K-factor~
 & $\quad \sigma_{1-loop} \quad$ & ~K-factor~  \\
\hline
LS2   & 5.589   & 4.545   & 0.813 & 5.867   & 1.050 \\
SPS1a & 0.04207 & 0.04145 & 0.985 & 0.04170 & 0.991 \\
\hline
\end{tabular}
\caption{Total cross sections (in pb) at Born and loop level and K-factors}
\label{tab:sigmaK}
\end{table}

It is possible to see that for both LS2 and SPS1a the corrections in the 2HDM subset are very small,
of the order of few percent, while in the complete SUSY case the light LS2 spectrum features a bigger correction
($\sim$19\%) than in the SPS1a case ($\sim$2\%).

\begin{figure}[t]
\centering
\epsfig{file=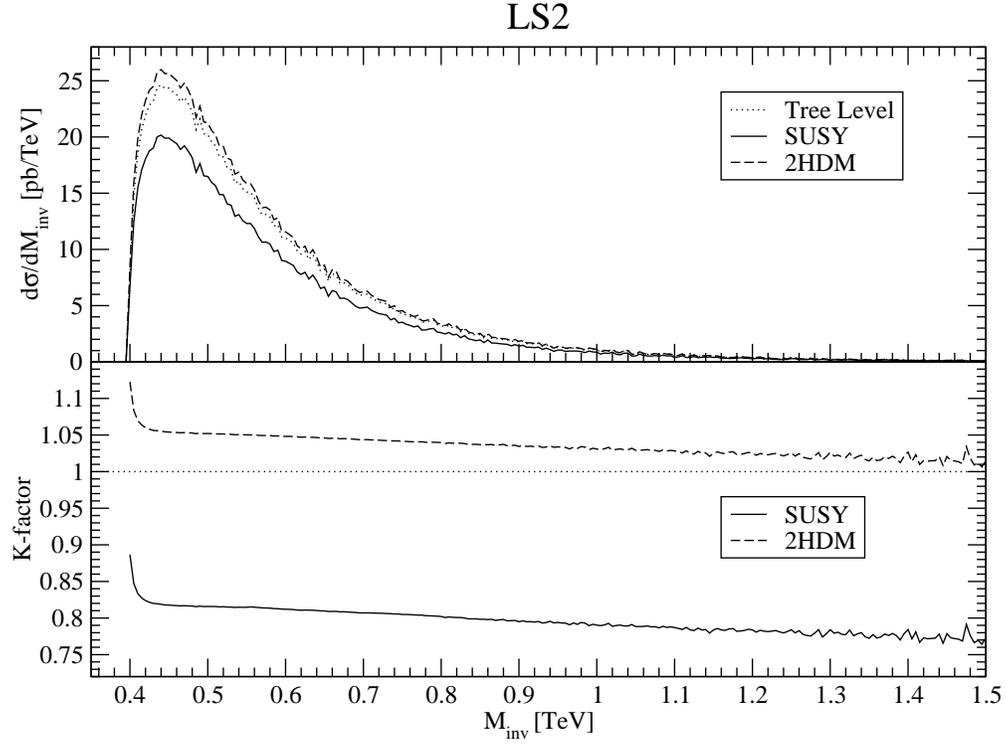, width=0.8\textwidth, angle=0}\\[35pt]
\epsfig{file=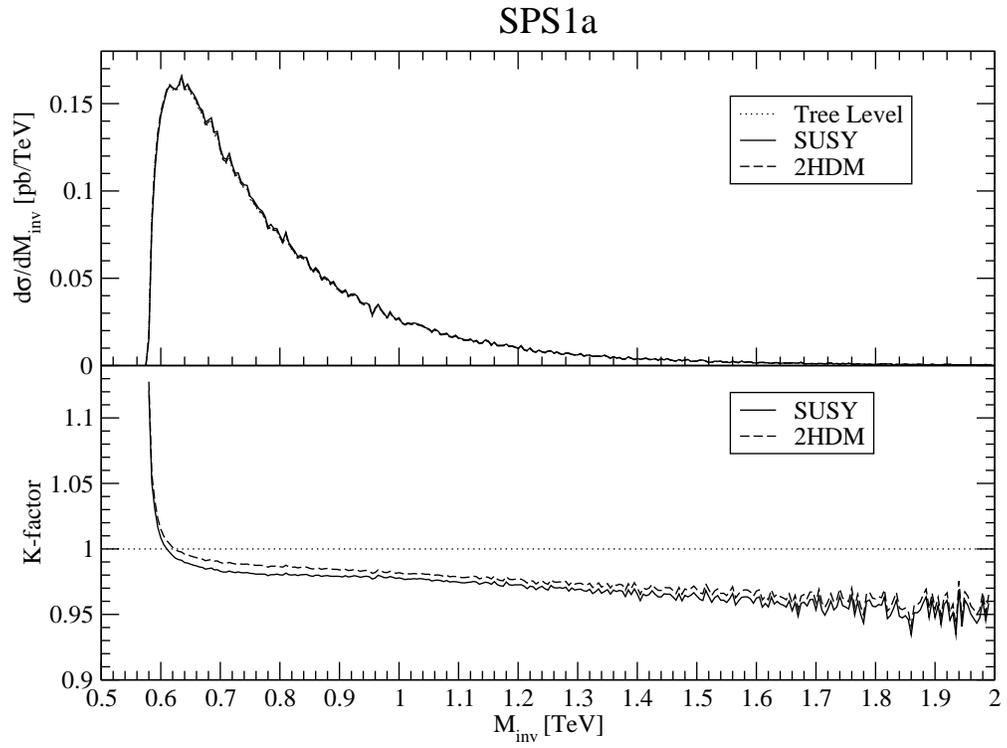, width=0.8\textwidth, angle=0}
\caption{Differential distribution (upper panels) and partial K-factors (lower panels) in LS2 and SPS1a.}
\label{fig:diffdistr}
\end{figure}

The differential distributions for the two benchmark points are shown in Fig.~\ref{fig:diffdistr},
where it is possible to see that, as a general behaviour,
the one-loop corrections decrease from the low invariant mass region to high energies.
In the SPS1a case (both SUSY and 2HDM) the one-loop corrections are positive near threshold,
but suddenly drop and become negative at high energies:
such compensating contributions are at the origin of the small one-loop correction to the total cross section.
In LS2, on the other hand, SUSY and 2HDM behave in different ways:
in the former case the one-loop corrections are always negative, the K-factor is
$\sim$0.97 near threshold and decreases at high energies with a behaviour analogous to the SPS1a case,
thus explaining the large negative correction to the total cross section in this scenario;
in the latter, the one-loop corrections are positive in a wider $M_{inv}$ range, giving rise to the positive overall
correction to the total cross-section.

\begin{figure}[t]
\centering
\epsfig{file=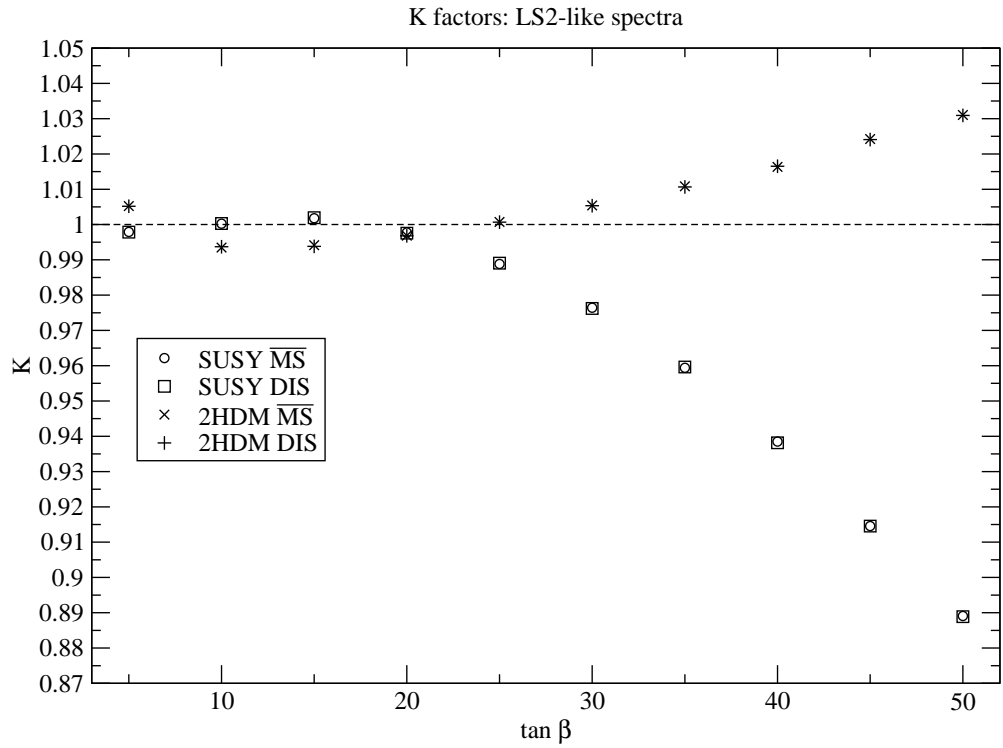, width=0.8\textwidth, angle=0}\\[35pt]
\epsfig{file=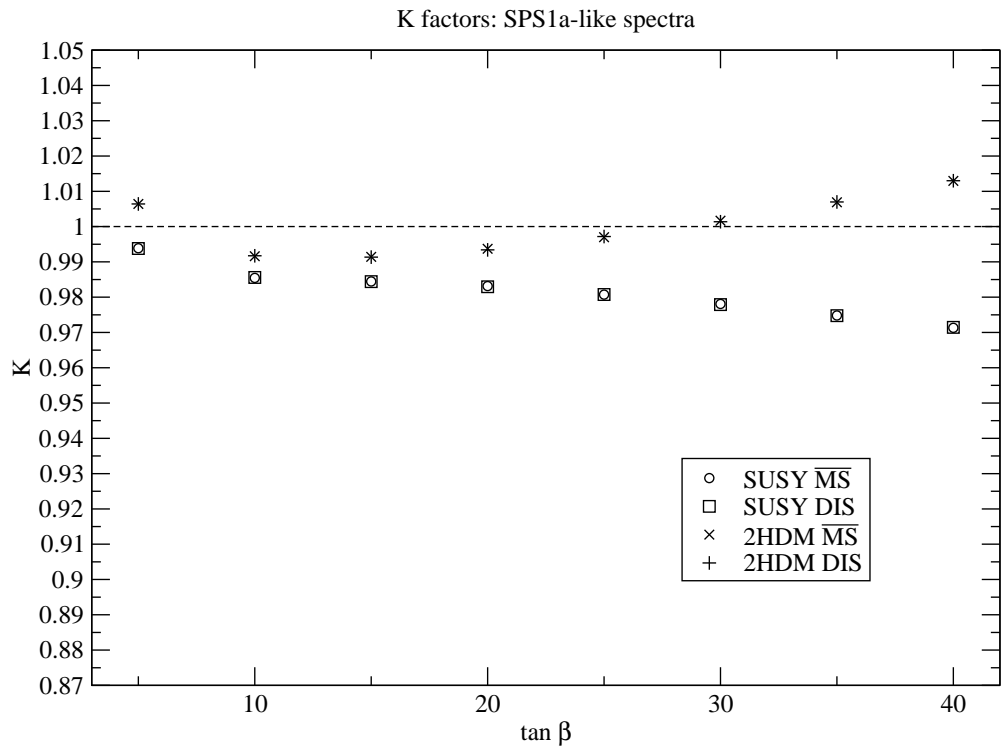, width=0.8\textwidth, angle=0}
\caption{K-factor dependence on $\tan\beta$ for LS2- and SPS1a-like spectra.}
\label{fig:tanbetadependence}
\end{figure}

As a second step in our analysis, given the relevance of $\tan\beta$ for the process under investigation,
we have also analyzed the dependence of the K-factors on this parameter.
Starting from the two previous LS2 and SPS1a spectra, we have considered $\tan\beta$ as a free parameter
and varied it at low energy within a reasonable range.
Chargino and neutralino masses and mixing matrices depend on the value of $\tan\beta$, and they have been 
varied accordingly.
The results of the K-factors as a function of $\tan\beta$ for the LS2- and SPS1a-like spectra
are shown in Fig.~\ref{fig:tanbetadependence}.
It is possible to notice that the dependence of the K-factor is stronger in the LS2 case: in the complete
SUSY scenario, it ranges from $\sim$1 (low $\tan\beta$) to $\sim$0.89 (large $\tan\beta$),
while the dependence in the 2HDM scenario shows opposite behaviour.
One sees that the NLO effects remain perturbatively under control in the whole considered range,
even for large values of $\tan\beta$, where corrections
are usually expected to become large.
Similar results have been found for the SPS1a-like spectra, where however, the dependence is milder than in the LS2-like case.

\begin{figure}[t]
\centering
\epsfig{file=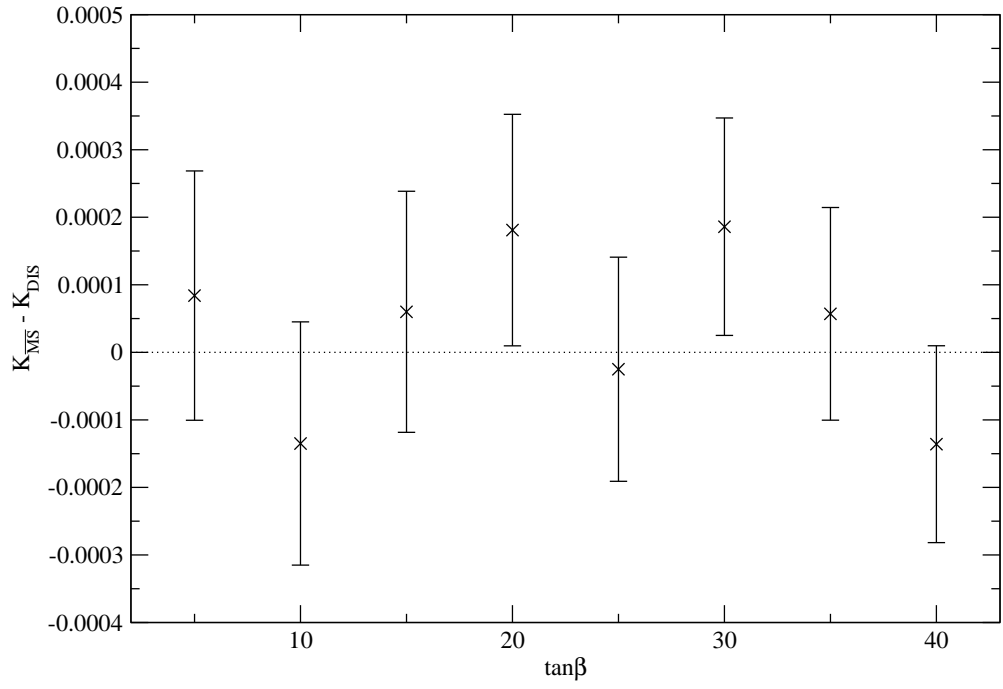, width=0.8\textwidth, angle=0}\\[35pt]
\caption{Factorization scheme dependence of the K-factor}
\label{fig:MSvsDIS}
\end{figure}

As a final remark, we can say that the dependence on the factorization scheme of our results is very mild, of the order of 0.01\%
in all the considered cases, and in Fig.~\ref{fig:MSvsDIS} the differences between the two schemes are shown in more detail.

\section{Conclusions}

In this paper we have calculated the complete EW NLO expression
of the $bg \to t H^-$ process both in a 2HDM and in the MSSM, assuming a mSUGRA symmetry breaking
scheme, to investigate the size of the corrections to tree-level observables and their $\tan\beta$ dependence.
In our calculation we have included the full computation of QED radiation, which
makes our analysis testable against future data.
We have considered two benchmark points characterized by quite different values of $\tan\beta$ (10 and 50) and
we have let the parameter vary into a reasonable range to investigate for dependences of the observables.
We have found that the NLO corrections to the total cross sections can be sizable (negative and of the order of 20\%)
in the LS2 point, which is characterized by a light spectrum,
and due to its cross sections ($\sim$5~pb) they might be hopefully observed at the LHC.

The dependence on $\tan\beta$ of the corrections is similar in the two benchmark points that we have analyzed,
but more enhanced in LS2. On the other hand, the corrections exhibit a different behaviour
in the two considered physical scenarios:
in the 2HDM the corrections are generally mild, of the order of a few relative percent in the whole scanned range,
and the effects raises for large $\tan\beta$; in the MSSM case, the one-loop corrections become negative and
decreasing for large values of $\tan\beta$.

Given the outcome of our computations, we conclude that a complete calculation of EW MSSM NLO effects is worth and
should be taken into account for a full, reliable and meaningful NLO analysis of this important process,
which is probably the only one that can provide information on the charged Higgs
couplings of the model.

\section*{Acknowledgements}

We would like to thank Edoardo Mirabella for his contribution to the calculation of QED radiation and
for valuable comments and suggestions.

\newpage

\appendix

\section{Counter terms and self-energy corrections to Born amplitudes}
\label{AppendixA}

In this appendix the expression of the countermterms are explicitely listed.
They concern the counterterms for $b$, $t$, $H^-$ lines
as well as the propagator self-energy corrections
for $b$ and $t$ exchanges.\\

\underline{s-channel counterterms}
 
\bqa
N^{c.t.~s}_{1L}=
&&-~{g_s({\lambda^l\over2})\over s-m^2_b}\{
{3\over2}\delta Z^b_{L}c^{L}(b\to t H^-)
+{1\over2}(\delta Z^t_{R}+\delta \psi_t)c^{L}(b\to t H^-)
\nonumber\\
&&
+\delta c^L(b\to t H^-)
+{1\over2}
\sum_{j}\delta Z^*_{j1}c^{L}(b\to t j)\}
\eqa
\bqa
N^{c.t.~s}_{1R}=
&&-~{g_s({\lambda^l\over2})\over s-m^2_b}\{
{3\over2}\delta Z^b_{R}c^{R}(b\to t H^-)
+{1\over2}(\delta Z^t_{L}+\delta \psi_t)c^{R}(b\to t H^-)
\nonumber\\
&&
+\delta c^R(b\to t H^-)
+{1\over2}
\sum_{j}\delta Z^*_{j1}c^{R}(b\to t j)\}
\eqa

\bqa
N^{c.t.~s}_{3L}={m_bg_s({\lambda^l\over2})\over s-m^2_b}
(\delta Z^b_{R}-\delta Z^b_{L})c^R(b\to t H^-)
\eqa
\bqa
N^{c.t.~s}_{3R}={m_bg_s({\lambda^l\over2})\over s-m^2_b}
(\delta Z^b_{L}-\delta Z^b_{R})c^L(b\to t H^-)
\eqa
\noindent
where, because of the $H^--G^-$ mixing, we denote $H^-$ by $j=1$ and $G^-$ by $j=2$.\\
And from $b$ s.e. one gets ($\eta=+1,-1$ refering to $R,L$ chiralities):
\bqa
N^{b~s.e.}_{1\eta}=
&&~g_s({\lambda^l\over2})
{c^{\eta}(b\to t H^-)\over(s-m^2_b)^2}
[s(\Sigma^b_{\eta}(s)+\delta Z^b_{\eta})
+m^2_b(\Sigma^b_{-\eta}(s)+\delta Z^b_{-\eta})
\nonumber\\
&&
+2m^2_b(\Sigma^b_{S}(s)-~{1\over2}
(\delta Z^b_{\eta}+
\delta Z^b_{-\eta})-~{\delta m_b\over m_b})]
\eqa

\bqa
N^{b~s.e.}_{3\eta}=
&&~g_s({\lambda^l\over2})
{c^{-\eta}(b\to t H^-)m_b\over(s-m^2_b)}
[\Sigma^b_{\eta}(s)+\delta Z^b_{\eta}
+\Sigma^b_{S}(s)\nonumber\\
&&-~{1\over2}
(\delta Z^b_{\eta}+\delta Z^b_{-\eta})-~{\delta m_b\over m_b}]
\eqa

\underline{u-channel counterterms}

\bqa
N^{ct~u}_{1L}&=&-g_s({\lambda^l\over2}){1\over(u-m^2_t)}
\{({3\over2}\delta Z^t_R+{1\over2}\delta \psi_t+
{1\over2}\delta Z^b_L)c^L(b\to t H^-)\nonumber\\
&&+\delta c^L(b\to t H^-)+{1\over2}
\sum_{j}\delta Z^*_{j1}c^L(b\to t j)\}
\eqa
\bqa
N^{ct~u}_{1R}&=&-g_s({\lambda^l\over2}){1\over(u-m^2_t)}
\{({3\over2}\delta Z^t_L+{1\over2}\delta \psi_t+
{1\over2}\delta Z^b_R)c^R(b\to t H^-)\nonumber\\
&&+\delta c^R(b\to t H^-)+{1\over2}
\sum_{j}\delta Z^*_{j1}c^R(b\to t j)\}
\eqa
\bqa
N^{ct~u}_{2L}&=&-2g_s({\lambda^l\over2}){1\over(u-m^2_t)}
\{({3\over2}\delta Z^t_R+{1\over2}\delta \psi_t+
{1\over2}\delta Z^b_L)c^L(b\to t H^-)\nonumber\\
&&+\delta c^L(b\to t H^-)+{1\over2}
\sum_{j}\delta Z^*_{j1}c^L(b\to t j)\}
\eqa
\bqa
N^{ct~u}_{2R}&=&-2g_s({\lambda^l\over2}){1\over(u-m^2_t)}
\{({3\over2}\delta Z^t_L+{1\over2}\delta \psi_t+
{1\over2}\delta Z^b_R)c^R(b\to t H^-)\nonumber\\
&&+\delta c^R(b\to t H^-)+{1\over2}
\sum_{j}\delta Z^*_{j1}c^R(b\to t j)\}
\eqa
\bqa
N^{ct~u}_{3L}&=&g_s({\lambda^l\over2}){m_tc^L(b\to t H^-)\over(u-m^2_t)}
\{\delta Z^t_R-\delta Z^t_L\}
\eqa
\bqa
N^{ct~u}_{3R}&=&g_s({\lambda^l\over2}){m_tc^R(b\to t H^-)\over(u-m^2_t)}
\{\delta Z^t_L-\delta Z^t_R\}
\eqa
and from $t$ s.e. one gets:
\bqa
N^{t~s.e.}_{1\eta}&=&g_s({\lambda^l\over2}){c_{\eta}(b\to t H^-)\over
(u-m^2_t)^2}[u(\Sigma^t_{-\eta}(u)+\delta Z^t_{-\eta})
+m^2_t(\Sigma^t_{\eta}(u)+\delta Z^t_{\eta})
\nonumber\\
&&+2m^2_t(\Sigma^t_{S}(u)-{1\over2}(\delta Z^t_{\eta}
+\delta Z^t_{-\eta})-{\delta m_t\over m_t})]
\eqa
\bqa
N^{t~s.e.}_{2\eta}&=&2N^{t~s.e.}_{1\eta}
\eqa
\bqa
N^{t~s.e.}_{3\eta}&=&
g_s({\lambda^l\over2}){c_{\eta}(b\to t H^-)m_t\over
(u-m^2_t)}[\Sigma^t_{\eta}(u)+\delta Z^t_{\eta}
+\Sigma^t_{S}(u)-{1\over2}(\delta Z^t_{\eta}
+\delta Z^t_{-\eta})-{\delta m_t\over m_t}]
\eqa

\noindent The c.t. appearing in the above expressions are obtained in terms of self-energies as follows:

{\bf \underline{b and t quark}}

\bq
\delta Z^b_L=\delta Z^t_L\equiv \delta Z_L=
-\Sigma^b_L(m^2_b)-m^2_b[\Sigma^{'b}_L(m^2_b)+\Sigma^{'b}_R(m^2_b)
+2\Sigma^{'b}_S(m^2_b)]
\eq
\bq
\delta Z^b_R=
-\Sigma^b_R(m^2_b)-m^2_b[\Sigma^{'b}_L(m^2_b)+\Sigma^{'b}_R(m^2_b)
+2\Sigma^{'b}_S(m^2_b)]
\eq
\bq
\delta Z^t_R=\delta Z_L+\Sigma^t_L(m^2_t)-\Sigma^t_R(m^2_t)
\eq
\bq
\delta\Psi_t=-\{\Sigma^t_L(m^2_t)+\delta Z_L+
m^2_t[\Sigma^{'t}_L(m^2_t)+\Sigma^{'t}_R(m^2_t)
+2\Sigma^{'t}_S(m^2_t)]\}
\eq
\bq
\delta m_b={m_b\over2}Re[\Sigma^b_L(m^2_b)+\Sigma^b_R(m^2_b)
+2\Sigma^b_S(m^2_b)]
\eq
\bq
\delta m_t={m_t\over2}Re[\Sigma^t_L(m^2_t)+\Sigma^t_R(m^2_t)
+2\Sigma^t_S(m^2_t)]
\eq

{\bf \underline{gauge boson}} 

\bq
\delta Z^W_1-\delta Z^W_2=~{\Sigma^{\gamma Z}(0)\over s_Wc_W M^2_Z}
\eq

\bq
\delta Z^W_2 = - \Sigma^{'\gamma\gamma}(0) 
+2{c_W\over s_W M^2_Z}\Sigma^{\gamma Z}(0) 
+{c^2_W\over s^2_W}[{\delta M^2_Z\over M^2_Z} - 
{\delta M^2_W\over M^2_W}]
\eq
\bq
\delta M^2_W=Re\Sigma^{WW}(M^2_W)~~~~~\delta M^2_Z=Re\Sigma^{ZZ}(M^2_Z)
\eq

{\bf \underline{Higgs boson}}\\

We need $\delta Z^*_{j1}$ which means
$\delta Z^*_{H^-H^-}$ and $Z^*_{G^-H^-}$. We use the on-shell procedure of 
Wan et al \cite{Wan:2001rt} in which
\bq
\delta Z_{H^-H^-}=-\Sigma^{'}_{H^-}(p^2=m^2_{H^-})
\eq
and
\bq
\delta Z^*_{G^-H^-}=\delta Z_{G^+H^+}=
-~{2\Sigma^*_{H^-W^-}(m^2_{H^-})\over
M_W}=
~{2\Sigma_{H^+W^+}(m^2_{H^+})\over
M_W}
\eq

{\bf \underline{Couplings}}\\

The Yukawa $btH^-$ coupling leads to the c.t.
$\delta c^L$ and $\delta c^R$, computed in terms
of $\delta g$, $\delta m_{t,b}$, $\delta M_W$
(given above)
and $\delta\tan\beta$. For the latter, we have adopted the renormalization scheme of~\cite{Wan:2001rt}.\\

\bq
{\delta c^L\over c^L}={\delta g\over g}+{\delta m_t\over m_t}
-{\delta M_W\over M_W}-{\delta\tan\beta\over\tan\beta}
\eq
\begin{equation}
\label{Yukct}
{\delta c^R\over c^R}={\delta g\over g}+{\delta m_b\over m_b}
-{\delta M_W\over M_W}+{\delta\tan\beta\over\tan\beta}
\end{equation}

\bq
{\delta g\over g}=\delta Z^W_1-{3\over2}\delta Z^W_2
\eq

\begin{equation}
\label{WHscheme}
{\delta\tan\beta\over\tan\beta}={Re\Sigma_{H^+W^+}(m^2_{H^+})
\over M_W\sin2\beta}
\end{equation}

\newpage

\hfill

\hfill

\hfill

\hfill

\hfill

\end{document}